\documentclass[journal,11pt,draftclsnofoot,onecolumn]{IEEEtran}

\ifCLASSINFOpdf

\else

\fi

\usepackage[T1]{fontenc}
\usepackage{flushend} 
\usepackage{upgreek}
\usepackage{stfloats}
\usepackage{subfig}
\usepackage{fancyhdr}
\usepackage{mathrsfs}
\usepackage{graphicx}
\usepackage{geometry}
\usepackage{color}
\usepackage{booktabs}
\usepackage{placeins}
\usepackage{float}
\usepackage{tabularx,colortbl}
\usepackage{diagbox} 
\usepackage{makecell,multirow}
\usepackage{multirow}
\usepackage{amsmath}
\usepackage{amsfonts}
\usepackage{amssymb}
\usepackage{amsthm}
\usepackage{bm}
\usepackage{algorithm}
\usepackage{algorithmicx}
\usepackage{algpseudocode}
\newcommand{\tabincell}[2]{\begin{tabular}{@{}#1@{}}#2\end{tabular}} 
\usepackage[numbers,sort&compress]{natbib}
\usepackage{etoolbox} 
\makeatletter
\patchcmd{\@makecaption}
{\scshape}
{}
{}
{}
\makeatother

\begin{document}

\title{Low-Complexity Channel Estimation for Extremely Large-Scale MIMO in Near Field}
\author{\IEEEauthorblockN{Chun~Huang, 
Jindan Xu\IEEEauthorrefmark{0}, \IEEEmembership{Member,~IEEE},
Wei~Xu\IEEEauthorrefmark{0}, \IEEEmembership{Senior Member,~IEEE}, \\
Xiaohu You\IEEEauthorrefmark{0}, \IEEEmembership{Fellow,~IEEE}, 
Chau Yuen\IEEEauthorrefmark{0}, \IEEEmembership{Fellow,~IEEE}, 
and Yijian~Chen\IEEEauthorrefmark{0}
}
\vspace{-1.1cm}
\thanks{
	
	Chun Huang, Wei Xu, and Xiaohu You are with the National Mobile Communications Research Laboratory (NCRL), Southeast University, Nanjing 210096, China (e-mail: huangchun@seu.edu.cn; wxu@seu.edu.cn; xhyu@seu.edu.cn).
	 
	Jindan Xu and Chau Yuen are with the School of Electrical and Electronics Engineering, Nanyang Technological University, Singapore 639798, Singapore   (e-mail: jindan1025@gmail.com; chau.yuen@ntu.edu.sg).
	
	
	

}
}

\maketitle
\thispagestyle{fancy}
\renewcommand{\headrulewidth}{0pt}
\pagestyle{fancy}
\cfoot{}
\rhead{\thepage}

\newtheorem{mylemma}{Lemma}
\newtheorem{mytheorem}{Theorem}
\newtheorem{mypro}{Proposition}
\begin{abstract}
\textbf{The extremely large-scale massive multiple-input multiple-output (XL-MIMO) has the potential to achieve boosted spectral efficiency and refined spatial resolution for future wireless networks.
However, channel estimation for XL-MIMO is challenging since the large number of antennas results in high computational complexity with the near-field effect. In this letter, we propose a low-complexity sequential angle-distance channel estimation (SADCE) method for near-field XL-MIMO systems equipped with uniformly planar arrays (UPA). 
Specifically, we first successfully decouple the angle and distance parameters, which allows us to devise a two-dimensional discrete Fourier transform (2D-DFT) method for angle parameters estimation.
Then, a low-complexity distance estimation method is proposed with a closed-form solution.
 Compared with existing methods, the proposed method achieves significant performance gain with noticeably reduced computational complexity.
Numerical results verify the superiority of the proposed near-field channel estimation algorithm.}
\end{abstract}


%
\IEEEpeerreviewmaketitle

\vspace{-0.5cm}
\section{Introduction}

\IEEEPARstart{E}{xtremely} large-scale massive multiple-input multiple-output (XL-MIMO) is a promising technology for the next-generation wireless networks, e.g., the sixth-generation (6G) \cite{NF1}-\cite{Xu2}. 
{\color{black}Compared with typical massive multiple-input multiple-output (MIMO) with 32 or 64 antennas, the number of antennas in an XL-MIMO systems is usually considered in the order-of-magnitude of $10^3$ or even more, which enables higher spectral efficiency and spatial resolution \cite{NF_Commun}.}
{\color{black}Note that the electromagnetic radiation field of near-field and far-field domains is separated by the classical Rayleigh distance $r_{\rm Rayl} = \frac{2D^2}{\lambda}$ \cite{Jin_gloabcom}}.
{\color{black}Due to the employment of extremely large-scale antenna arrays in XL-MIMO systems, the Rayleigh distance extends up to several hundred meters, making the near-field domain more probable in practice. In particular, for a typical setup of 256-antenna MIMO at 10 GHz, the Rayleigh distance reaches around 245.8 meters.} 

For near-field propagation, the conventional far-field channel estimation methods are no longer suitable for near-field communication systems due to the fact that a near-field channel is characterized by two parameters \cite{RD-MUSIC-ULA}, i.e., the angle-of-departure/arrival (AoD/AoA) and the transmission distance between users and the base station (BS). 
{\color{black}To facilitate the acquisition of channel state information (CSI) parameters of near-field sources, classic high resolution subspace-based parameter estimation algorithms, such as the multiple signal classification (MUSIC) algorithm \cite{2D-MUSIC}, have attracted enormous interests due to its high resolution estimation. To address the expensive computational costs associated with conventional MUSIC algorithm, a low-complexity algorithm based on reduced-dimension (RD) MUSIC algorithm was proposed in \cite{RD-MUSIC-ULA}. However, the the proposed method  still required a significant amount of matrix operations, such as matrix decomposition, which is not suitable for the XL-MIMO systems with an excessive number of antennas.} 
Besides, conventional far-field sparse channel estimation methods have recently been extended to near-field domain in an XL-MIMO system, e.g., codebook-based beam training \cite{Dai_cb2}--\cite{You_Fast_BT}. Specifically, the authors in \cite{Dai_cb2} designed a polar-domain codebook of the near-field channel based on the Fresnel approximation. 
However, this method required a two-dimensional exhaustive search for all possible beam directions and distances, which resulted in high training overhead and increased computational complexity. 
To further reduce the training overhead, a two-stage beam training method was proposed in \cite{You_Fast_BT}, which decomposed the two-dimensional search into two sequential phases. 
Nevertheless, beam training methods still required a significant number of pilots to sequentially sweep these codewords.

In practice, considering the extremely large number of antennas in XL-MIMO systems, uniform planar array (UPA), rather than uniformly linear array (ULA), is a more feasible compact antenna array deployment, which can also avoid the beam squint in ULA systems \cite{UPA_good}. 
However, the above researches, e.g., \cite{Dai_cb2}--\cite{You_Fast_BT}, focused on the ULA case and do not readily extend to the UPA case \cite{Jin_arxiv}, especially in the near field. Specifically, the number of angle parameters doubles that of a UPA system and all these parameters couple to the distance parameter. 
As such, a pragmatic low-complexity channel estimation method is expected for XL-MIMO to avoid high complexity of a three-dimensional grid search of parameters, i.e., azimuth angles, elevation angles, and distance.

{\color{black}
	Motivated by the aforementioned challenges, we investigate the channel estimation problem in the near field of an XL-MIMO system with UPA in this letter. A novel sequential angle-distance channel estimation (SADCE) algorithm is presented to estimate the near-field channels for XL-MIMO systems with UPA. This algorithm successfully decouples the angles and distance in the channel estimation problem and enables a low-complexity solution with partially closed-form calculations.   
		Simulation results verify the effectiveness of the proposed SADCE algorithm. Compared with existing state-of-the-art algorithms, the proposed method shows superiority in terms of not only the estimation accuracy but also the computational complexity.           
	
}

\emph{Notations}: Throughout this paper, superscripts ${(\cdot)}^{\rm H}$, ${(\cdot)}^{\rm T}$, ${(\cdot)}^*$, and ${(\cdot)}^{-1}$ denote the conjugate transpose, transpose, conjugate, and inverse of the input entity, respectively.
Operators ${\|\mathbf{\cdot}\|}$ and ${\|\mathbf{\cdot}\|}_{\rm F}$ stand for the Euclidean norm of a vector and the Frobenius norm of a matrix. 
$\mathbb{E}\{\cdot\}$ represents the expectation of the input
entity. ${\rm Diag}({\bf x})$ represents the diagonal matrix $\bf X$ obtained from vector $\bf x$.
Denote by $\mathcal{CN}({\mu},{\sigma^2})$ as the distribution of a circularly symmetric complex Gaussian variable with mean ${\mu}$ and variance ${\sigma^2}$. 
The $k$th entry of vector $\mathbf x$ and the $(i,j)$th element of matrix $\mathbf{X}$ are represented by $[\mathbf  x]_{k}$ and $[\mathbf{X}]_{ij}$, respectively. 
Operator $\frac{\partial(\cdot )}{\partial \mathbf x}$ stands for the first-order derivative of the input entity with respect to $\mathbf x$.


\vspace{-0.3cm}
\section{System Model}
\vspace{-0.1cm}

\begin{figure}[t]
	\centering
	\vspace{-0.5cm}
	\includegraphics[width =2.5in]{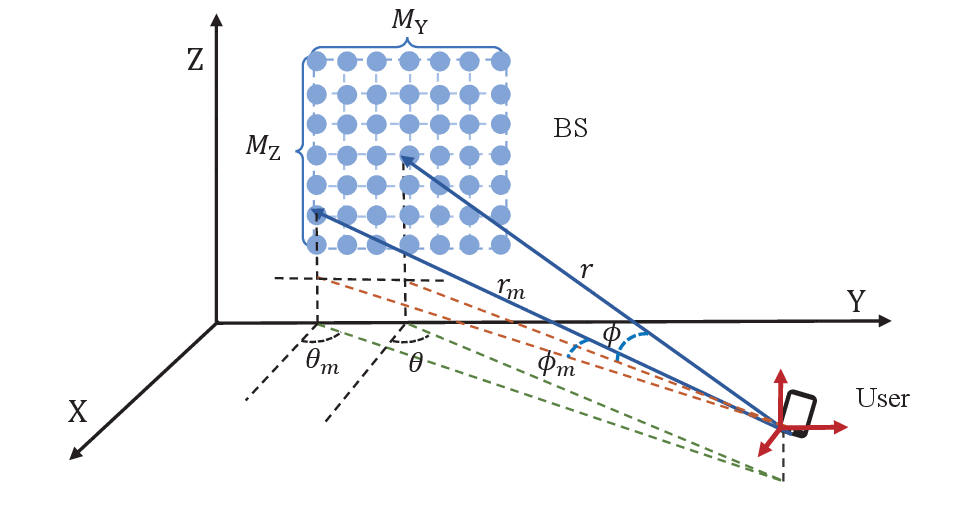}
	\vspace{-0.2cm}
	\caption{An XL-MIMO system in near-field domain.}
	\label{Fig:P_system}
	\vspace{-0.6cm}
	
\end{figure}

We consider the uplink of an XL-MIMO system, where the BS is located in the YOZ plane to serve multiple single-antenna users, as shown in Fig. \ref{Fig:P_system}. The BS is equipped with an $M$-antenna  
UPA, where the number of antennas along the Y-axis and Z-axis are represented by $M_{\rm Y}$ and $M_{\rm Z}$, respectively. 
Without loss of generality, we consider a near-field single-antenna user.
Due to the severe path loss effect in mmWave/Thz frequency, the channel gains of the non-line-of-sight (NLoS) path are negligible compared to that of line-of-sight (LoS) path.
As commonly adopted in literature, e.g., \cite{RD-MUSIC-ULA}, \cite{Dai_cb2}, and \cite{RIS_CE1}, we consider the LoS path and the near-field channel model between the BS and the user as 
\begin{align}\label{eq:h1}
\mathbf{h} = \beta {\mathbf b}(\theta,\phi,r),
\end{align}
where parameters $\beta$, $\theta$, $\phi$, and $r$ denote the complex channel gain, azimuth angle, elevation angle, and the distance between the user and the BS, respectively, and ${\mathbf b}(\theta,\phi,r)$ is the near-field steering vector  
\begin{align}\label{eq:b_vector}
	 {\mathbf b}(\theta,\phi,r) = \frac{1}{\sqrt{M}}\left[{\rm e}^{-\frac{j2\pi}{\lambda}(r_0-r)},...,{\rm e}^{-\frac{j2\pi}{\lambda}(r_{M-1}-r)}\right] ,
\end{align}
where $r_m$ denotes the distance from the $m$th antenna of the BS to the user, and $\lambda$ denotes the wavelength. The index of the antennas is denoted by $m = M_{\rm Y} m_z+m_y+M/2$, where $m_y=-(M_{\rm Y}-1)/2,...,(M_{\rm Y}-1)/2$ and $m_z=-(M_{\rm Z}-1)/2,...,(M_{\rm Z}-1)/2$. $M_{\rm Y}$ and $M_{\rm Z}$ are assumed to be odd numbers. 
For notational simplicity, define $u \triangleq {\rm sin}\theta$ and $v \triangleq {\rm cos}\theta{\rm sin}\phi$. {\color{black} Utilizing a two-dimensional Fresnel approximation \cite{Jin_gloabcom},\cite{Dai_cb2}, the array response can be simplified  as  }
\begin{align}\label{eq:b2}
	[{\mathbf b}(u,v,r)]_m = {\rm e}^{-\frac{j2\pi}{\lambda}(p_m(u,v)-q_m(u,v,r))},
\end{align}
where $p_m(u,v)=m_zud-m_yvd$, $q_m(u,v,r)= \frac{1}{2r}\left(m_z^2d^2+m_y^2d^2-(m_zud-m_yvd)^2 \right)$, and $d$ is the spacing between neighbor antennas. To avoid ambiguity, the antenna spacing can be chosen as $d = \lambda/4$ \cite{Jin_gloabcom}--\cite{RD-MUSIC-ULA}.

During the period of uplink transmission, the user sends a sequence of $L$ symbols to the BS.
Then, at time instant $l\in\{1,\cdots,L\}$, the received signal at BS can be expressed as
\begin{align}\label{eq:yl}
\mathbf{y}_l={\mathbf h}{{s}_l}+{\mathbf{z}_l}=\beta{\bf b}(u,v,r){{s}_l}+{\mathbf{z}_l},
\end{align}
where $s_l\in \mathbb{C}$ stands for the signal transmitted by the user with a power constraint $\mathbb{E}\{\vert s_l \vert^2\}=P$, and ${\mathbf z}_l \sim {\mathcal{CN}}(0,\sigma^2{\mathbf{I}}_{ M})\in {\mathbb{C}}^{M\times 1}$ denotes the additive white Gaussian noise (AWGN) at the BS.

\vspace{-0.4cm}
\section{Proposed Near-Field Channel Estimation}
\vspace{-0.1cm}

In this section, we propose a novel channel estimation protocol for the XL-MIMO system. 
The main idea is to enable sequential estimation by decoupling the angles ${u, v}$ and distance $r$, and then reconstruct the channel between the user and BS.
To obtain an initialized channel estimate, we adopt the typical least square (LS) method \cite{Xu4} to get 
\begin{align}\label{eq:h_LS}
	\hat{\bf h} = {\bf Y}{\bf s}^{*}({\bf s}^{\rm H}{\bf s})^{-1},
\end{align}
where ${\bf s} = [s_1,...,s_L]^{\rm T} \in {\mathbb C}^{L\times 1} $ denotes the pilots transmitted by the user, and ${\mathbf Y} = [{\bf y}_1,...,{\bf y}_L]\in {\mathbb C}^{M\times L}$ denotes the received signals at the BS. 
Then, an estimate of the array covariance matrix can be expressed as ${\hat{\mathbf R} }={\hat {\bf h}}{\hat{\bf h} }^{\rm H}$. In the ideal noiseless case, the true covariance matrix $\bf R$ is from \eqref{eq:h1} as
\begin{align}\label{eq:R1}
	{\mathbf R}={\bf h}{\bf h}^{\rm H}=\vert \beta\vert^{2}{\bf b}(u,v,r){\bf b}^{\rm H}(u,v,r).
\end{align}
From \eqref{eq:b2}, the ($i$, $j$)th element of ${\bf R}\in {\mathbb{C}}^{M\times M}$ is
\begin{align}\label{eq:Rij}\nonumber
[\textbf{R}]_{i,j}&=\vert\beta\vert^2[{\bf b}(u,v,r)]_{i}[{\bf b}^{\rm H}(u,v,r)]_{j}\\ 
&=\vert\beta\vert^2{\rm e}^{-\frac{j2\pi}{\lambda}\left(p_i\left(u,v\right)-p_j\left(u,v\right) +q_i\left(u,v,r\right)-q_j\left(u,v,r\right) \right)} .
\end{align}	

{\color{black}After obtaining the covariance matrix $\bf R$, a pair of angles and distance can be directly estimated by performing a 3D MUSIC-like spectrum spectral search over the angle and distance domains.} However, the conventional MUSIC algorithm incurs tremendously expensive computational costs, which is usually hard to implement \cite{Jin_arxiv}. 
To circumvent this difficulty, we propose a method where the distance and angle parameters are decoupled and then a computational-efficient channel estimation method is developed in the following.
\vspace{-0.5cm}
\subsection {The Angle Parameter Estimation Algorithm}
\emph{1) Initial Angle Parameter Estimation.} 
{\color{black} To reduce the complexity, we propose an approach that the distance and angular parameters are decoupled and then estimated sequentially.}  
 {\color{black}Note that the anti-diagonal elements of $\bf R$ depend on only the angle parameters $(u,v)$, which is given by
\begin{align} 
[\mathbf{R}]_{i,M-i}\overset{({\rm a})}{=}{\rm e}^{-\frac{j2\pi}{\lambda}(2p_i(u,v))} 
={\rm e}^{j\frac{2\pi}{\lambda}\left(2i_zud-2i_yvd\right)},
\end{align}
where the equality in $(\rm a)$ uses  $q_i\left(u,v,r\right)-q_{M-i}\left(u,v,r\right)= 0$ and $p_i\left(u,v\right)-p_{M-i}(u,v) = 2p_i(u,v)$.} Collect all the anti-diagonal elements from $\bf R$, and then establish an updated covariance matrix ${\bf R}_a\in C^{M_{\rm Y}\times M_{\rm Z}}$ as follows
\begin{flalign} \label{eq:Ra_1}
\begin{tiny}		
		 {\bf R}_{a}=  
	     \left[
		\setlength{\arraycolsep}{0.18pt}
		\begin{array}{ccc}
			{ \tiny {\rm e}^{\frac{j 2 \pi}{\lambda}\left(-\left(M_{\rm Y}-1\right) v d+\left(M_{\rm Z}-1\right) u d\right)}} & \cdots & {\rm e}^{\frac{j 2 \pi}{\lambda}\left(-\left(M_{\rm Y}-1\right) v d-\left(M_{\rm Z}-1\right) u d\right)} \\
			\vdots & \ddots & \vdots \\
			{\rm e}^{\frac{j 2 \pi}{\lambda}\left(\left(M_{\rm Y}-1\right) v d+\left(M_{\rm Z}-1\right) u d\right)} & \cdots & {\rm e}^{\frac{j 2 \pi}{\lambda}\left(\left(M_{\rm Y}-1\right) v d-\left(M_{\rm Z}-1\right) u d\right)}
		\end{array}\right].
\end{tiny}	
\end{flalign}
Specifically, the ($i,j$)th element of $\mathbf{R}_{a}$ equals
\begin{align}
		[\mathbf{R}_a]_{i,j}={\rm e}^{\frac{j2\pi}{\lambda}\left(2ivd-(M_{\rm Y}-1)vd-2jud+(M_{\rm Z}-1)ud\right)},
\end{align}
where $a=0,...,M_{\rm Y}-1$ and $b=0,...,M_{\rm Z}-1$. 
Fortunately, after completing these steps, the angle parameters and distance are successfully decoupled. 
Therefore, we can modify the DFT-method in \cite{Gao_FF_DFT_UPA} to estimate the angle parameters in the near field. 

Defining two DFT matrices ${\mathbf U}_{\rm Y}\in{\mathbb C}^{M_{\rm Y}\times M_{\rm Y}}$ and ${\mathbf U}_{\rm Z}\in{\mathbb C}^{M_{\rm Z}\times M_{\rm Z}}$, we apply the DFT transformation of ${\mathbf R}_a$ to yield
\begin{small} 
	\begin{align}\label{eq:Ra_DFT}\nonumber
		&[{\bf R}_{a, \mathrm{DFT}}]_{i_{y}, i_{z}}\\ \nonumber
		&= [{\mathbf U}_{\rm Y} {\mathbf R}_a {\mathbf U}_Z]_{i_{y}, i_{z}}\\ \nonumber
		&=\frac{1}{M_{\rm Y} M_{\rm Z}} \sum_{m_{z}=0}^{\left(M_{\rm Z}-1\right)} \sum_{m_{y}=0}^{\left(M_{\rm Y}-1\right)} [{\bf R}_{a}]_{m_{y}, m_{z}} \times {\rm e}^{-j 2 \pi\left(\frac{m_{z} i_{z}}{M_{\rm Z}}+\frac{m_{y} i_{y}}{M_{\rm Y}}\right)}\\ 
		&\triangleq\frac{1}{M_{\rm Y} M_{\rm Z}} W_{v}(v,i_{y}) W_{u}(u,i_{z}),
	\end{align}
\end{small}%
where $W_{v}(v,i_{y})$ is defined as a binary function of $v$ and $i_y$, and  so is $W_{u}(u,i_{z})$ as a binary function of $u$ and $i_z$, which are respectively given by
\begin{small}
\begin{align}
	W_{v}(v,i_{y}) = \frac{\sin \left(\pi M_{\rm Y}\left(\frac{2 v d}{\lambda}-\frac{i_{y}}{M_{\rm Y}}\right)\right)}{\sin \left(\pi\left(\frac{2 v d}{\lambda}-\frac{i_{y}}{M_{\rm Y}}\right)\right)} {\rm e}^{j \pi\left(M_{\rm Y}-1\right)\left(\frac{2 v d}{\lambda}-\frac{i_{y}}{M_{\rm Y}}\right)}, 
\end{align}	
\end{small}%
\vspace{-0.2cm}%
and
\vspace{-0.2cm}
\begin{small}
	\begin{align}
W_{u}(u,i_{z}) = \frac{\sin\left(\pi M_{\rm Z}\left(\frac{2u d}{\lambda}+\frac{i_{z}}{M_{\rm Z}}\right)\right)}{\sin\left(\pi\left(\frac{2u d}{\lambda}+\frac{{i}_{z}}{M_{\rm Z}}\right)\right)}{\rm e}^{-j\pi\left(M_{\rm Z}-1\right)\left(\frac{2ud}{{\lambda}}+\frac{{i}_{z}}{M_{\rm Z}}\right)}.
	\end{align}	
\end{small}%
Applying the L'Hospital rule, it is straightforward to check that $W_{v}(v,i_{y})$ and $W_{u}(u,i_{z})$ reach their maximum values when both $\left(\frac{2 v d}{\lambda}-\frac{i_{y}}{M_{\rm Y}}\right)$ and $\left(\frac{2u d}{\lambda}+\frac{i_{z}}{M_{\rm Z}}\right)$ tend to zero.
Additionally, when the BS has an infinite number of antennas, i.e. $M_{\rm Y} \rightarrow \infty,M_{\rm Z}\rightarrow \infty$,  there must exist an integer pair $(\overline {i}_y,\overline{i}_z)$ where $[{\bf R}_{\rm a,DFT}]_{\overline {i}_y,\overline{i}_z}$ reaches its peak value, while the other elements of ${\bf R}_{\rm a,DFT}$ tend to zero.
Thus, an initial estimation of angle parameters can be chosen as
\begin{small}
	\begin{align}\label{angle_initial}
	{\hat u}=-\frac{\lambda \overline{i}_z}{2dM_{\rm Z}},\quad {\hat v}=\frac{\lambda \overline i_y}{2dM_{\rm Y}}.
	\end{align}
\end{small}
\emph{2) Bias Compensation through Angle Rotation.}
In practice, the finite values of $M_{\rm Y}$ and $M_{\rm Z}$ may lead to the power leakage problem \cite{PCH_RIS_DFT_UPA}-\cite{Xu3}. The resolution of the estimated angle parameters $(u,v)$ in \eqref{angle_initial} is constrained by $\frac{\lambda }{2dM_{\rm Z}}$ and  $\frac{\lambda}{2dM_{\rm Y}}$. In this section, an angle rotation operation is adopted to improve the estimation accuracy of the angle parameters.

To compensate the estimated AoAs, we define two rotation matrices ${\bf R}_{\rm Y}\left(M_{\rm Y},\Delta_{v}\right)$ and ${\bf R}_{\rm Z}\left(M_{\rm Z},\Delta_{u}\right)$ as 
\begin{align}
	{\bf R}_{\rm Y}\left(M_{\rm Y},\Delta_{v}\right) & =\operatorname{Diag}\left(1, {\rm e}^{j \Delta_{v}}, \ldots, {\rm e}^{j\left(M_{\rm Y}-1\right) \Delta_{v}}\right),
\end{align}
and 
\begin{align}
	{\bf R}_{\rm Z}\left(M_{\rm Z},\Delta_{u}\right) & =\operatorname{Diag}\left(1, {\rm e}^{j \Delta_{u}}, \ldots, {\rm e}^{j\left(M_{\rm Y}-1\right) \Delta_{u}}\right),
\end{align}
where $\Delta_{v}\in[-\frac{\lambda\pi}{dM_{\rm Y}}$  $\frac{\lambda\pi}{dM_{\rm Y}}]$  and $\Delta_{u}\in[-\frac{\lambda\pi}{dM_{\rm Z}},\frac{\lambda\pi}{dM_{\rm Z}}]$. The specific angle rotation calculation process of $[{\bf R}_{a, \mathrm{DFT}}]_{\overline i_{y}, \overline i_{z}}$ is then represented as 
\begin{small}
	\begin{flalign}\label{R_fine} \nonumber 
		&{[{\bf R}_{a}(\Delta_u,\Delta_{v})]_{\overline i_{y}, \overline i_{z}}} \\ \nonumber
		&= [{\mathbf U}_{\rm Y}]_{\overline{i}_{y},:} {\bf R}_{\rm Y}\left(M_{\rm Y},\Delta_{v}\right){\mathbf R}_a {\bf R}_{\rm Z}\left(M_{\rm Z},\Delta_{u}\right) [{\mathbf U}_{\rm Z}]_{:,\overline{i}_z} \\ 
		&=\frac{1}{M_{\rm Y} M_{\rm Z}} W_{v}(v+\frac{\lambda\Delta_v}{4\pi d},\overline i_{y}) W_{u}(u-\frac{\lambda\Delta_u}{4\pi d},\overline i_{z}),    
	\end{flalign}
\end{small}%
Through the angle rotation, the power becomes highly concentrated around the frequency point $(\overline i_{y},\overline i_{z})$. {\color{black}Then, the optimal rotation parameters $(\Delta_u,\Delta_{v})$ in \eqref{R_fine} can be obtained via a 2-D search within a small region as follows}
\begin{small}
	\begin{align} \label{eq:a_r_max}
	\left(\widehat{\Delta}_u,\widehat{\Delta}_{v}\right)= \underset{{\Delta_{u}\in[-\frac{\lambda\pi}{dM_{\rm Z}},\frac{\lambda\pi}{dM_{\rm Z}}], \Delta_{v}\in[-\frac{\lambda\pi}{dM_{\rm Y}},\frac{\lambda\pi}{dM_{\rm Y}}]}}{\arg \max} {[{\bf R}_{a}(\Delta_u,\Delta_{v})]_{\overline i_{y}, \overline i_{z}}}.
\end{align}
\end{small}%
Based on the derived optimal rotation parameters in \eqref{eq:a_r_max}, the accurate estimated angle parameters in \eqref{angle_initial} are refined as
\begin{small}
	\begin{align}\label{angle_fine}
	{\hat u}=\frac{\lambda}{2 d}\left(\frac{\widehat \Delta_{u}}{2 \pi}-\frac{\overline{i}_z}{M_{\rm Z}}\right),
	\; {\hat v}=\frac{\lambda}{2 d}\left(\frac{\overline i_{y}}{M_{\rm Y}}-\frac{\widehat \Delta_{v}}{2 \pi}\right).
	\end{align} 
\end{small}%
{\color{black} Then, the angle resolutions after compensation are respectively optimized to  $\frac{\lambda }{2dM_{\rm Z}G_{\rm Z}}$ and  $\frac{\lambda}{2dM_{\rm Y}G_{\rm Z}}$, which represent a $G_{\rm Z}$ and a $G_{\rm Y}$ fold enhancement compared to the initialization in \eqref{angle_initial}}

\vspace{-0.3cm}
\subsection {The Distance Estimation}
{\color{black}After obtaining the estimates of angles, the distance can be estimated by conducting one-dimensional MUSIC-based spectral searching within the near-field domain. 
	However, these methods require a significant amount of computational complexity and searching time, and the distance estimates are subject to gridding limition.
To tackle this issue, we derive a low-complexity distance estimation method by splitting the near-field steering vector and reconstructing the conventional MUSIC spectrum function in \cite{2D-MUSIC}.}

 Specifically, the near-field steering vector in \eqref{eq:b_vector} is split in terms of the angle parameters and distance as follows 
\begin{small}
	\begin{align} \label{D_Tt}\nonumber
		&\mathbf{b}(u, v, r) \\			   \nonumber
		&=\mathbf{Q}(u, v) \mathbf{t}(u, v, r) 	\\
		&=\underbrace{\left[
			\setlength{\arraycolsep}{0.2pt}
			\begin{array}{lll}
				\setlength{\arraycolsep}{0.2pt}	
				\mathrm{e}^{-\frac{j2\pi}{\lambda} p_{0}(u, v)} &  & \\
				& ...  & \\
				& &  \mathrm{e}^{- \frac{j2 \pi}{\lambda} p_{M-1}(u, v)}
			\end{array}\right]}_{\mathbf{Q}(u, v)}
	\underbrace{\begin{bmatrix} 
			\mathrm{e}^{-\frac{j 2 \pi}{\lambda} q_{0}(u, v, r)} \\
			\ldots \\
			\mathrm{e}^{- \frac{j2 \pi}{\lambda} q_{M-1}(u, v, r)}
	\end{bmatrix}}_{\mathbf{t}(u, v, r)},
	\\ \nonumber
	\end{align}
\end{small}%
where ${\bf Q}(u,v)\in {\mathbb C}^{M\times M} $ contains only the parameters of azimuth and vertical angles, and ${\bf t}(u,v,r)$ contains both the angle parameters and distances. 

{\color{black}
	Utilizing the orthogonality between the signal subspace and the noise subspace, the channel estimation problem can be addressed by minimizing the following 3D-MUSIC spectrum function
	\begin{align} \label{eq:f3D} 
	{\rm f}_{\rm MUSIC}(u, v, r)=\frac{1}{\mathbf{b}(u, v, r)^{\mathrm{H}} \mathbf{U}_{n} \mathbf{U}_{n}^{\mathrm{H}} \mathbf{b}(u, v, r)}. 
	\end{align}
	Combining \eqref{D_Tt} and \eqref{eq:f3D}, we can reconstruct the  spectrum function as 
	\begin{small}
		\begin{align} \label{eq:V1} \nonumber
		{\rm V}_{\rm MUSIC}(u, v, r) &=\frac{1}{{\rm f}_{\rm MUSIC}(u, v, r)} \\ \nonumber
		& =\mathbf{t}(u, v, r)^{\mathrm{H}} \mathbf{Q}(u, v)^{\mathrm{H}} \mathbf{U}_{n} \mathbf{U}_{n}^{\mathrm{H}} \mathbf{Q}(u, v) \mathbf{t}(u, v, r) \\
		& =\mathbf{t}(u, v, r)^{\mathrm{H}} \mathbf{T}(u, v) \mathbf{t}(u, v, r),
		\end{align}
	\end{small}%
}where $\mathbf{U}_{n}$ is the noise subspace and $\mathbf{T}(u, v) = \mathbf{Q}(u, v)^{\mathrm{H}} \mathbf{U}_{n} \mathbf{U}_{n}^{\mathrm{H}} \mathbf{Q}(u, v)$. Consequently, the channel parameters are estimated by solving the following optimization problem as
\begin{small}
	\begin{align} \label{eq:problem2} \nonumber
	&\min\limits_{u, v, r}\ \, \mathbf{t}(u, v, d)^{\mathrm{H}} \mathbf{T}(u, v) \mathbf{t}(u, v, r)^{\mathrm{H}} \\ 
	&\text { s.t. } \; \; \, {\bf c}^{\mathrm{H}} \mathbf{t}(u, v, r)=1,
	\end{align}
\end{small}%
where ${\bf c} = [\underbrace{0,...,0}_{{M-1}/{2}},1,\underbrace{0,...,0}_{{M-1}/{2}}]^{\rm T}\in{\mathbb C}^{M\times 1}$. 
Define the following Lagrange function of \eqref{eq:problem2} as
\begin{small}
	\begin{align}\label{eq:L}
		\mathcal{L}=\mathbf{t}(u, v, r)^{\mathrm{H}} \mathbf{T}(u, v) \mathbf{t}(u, v, r)^{\rm H}-\epsilon\left({\bf c}^{\mathrm{H}} \mathbf{t}(u, v, r)-1\right),
	\end{align}
\end{small}%
where $\epsilon$ is the Lagrange multiplier. Then, we minimize the objective function in \eqref{eq:problem2} by setting the following partial derivative of $\mathcal{L}$ in \eqref{eq:L} w.r.t. $\mathbf{t}(u, v, r)$ to zero. It gives 
\begin{align}
	\label{L_partial} \frac{\partial \mathcal{L}}{\partial \mathbf{t}(u, v, r)} & =2 \mathbf{T}(u, v) \mathbf{t}(u, v, r)+\epsilon \mathbf{c}=0, \\
	\label{t_esti}\mathbf{t}(u, v, r) & =\frac{\mathbf{T}(u, v)^{-1} \mathbf{c}}{\mathbf{c}^{\mathrm{H}} \mathbf{T}(u, v)^{-1} \mathbf{c}}.
\end{align}

Substituting the angle estimates of $\hat u$ and $\hat v$ into \eqref{L_partial}, 
we can obtain ${\hat{\mathbf t} }({\hat u}, {\hat v}, r)$. 
{\color{black} According to the definition in \eqref{D_Tt}, the phase of ${\mathbf{t}}(u, v, r)$ can be expressed by}
\begin{align} \nonumber
\hat{\mathbf{a}}  =\angle  {\mathbf{t}}(\hat u, \hat v, r) =\frac{1}{r} \times{{\bf f}(\hat u, \hat v)},
\end{align}
where
\begin{small}
	\begin{align}
	[{\bf f}(\hat u, \hat v)]_m=-\frac{\pi d^{2}}{\lambda}\left(\left(m_{z}\right)^{2}+\left(m_{y}\right)^{2}-\left(m_{z} {\hat u}-m_{y} {\hat v}\right)^{2}\right).
\end{align} 
\end{small}%
\begin{algorithm}[t]
	\caption{\color{black}The Proposed Near-Field Sequential Angle-Distance Channel Estimation (SADCE) algorithm} 
	\hspace*{0.02in} {\bf Input:} 
	${\mathbf Y}$, $\bf s$.\\
	\hspace*{0.02in} {\bf Output:} 
	${\hat {\mathbf h}}$.
	\begin{algorithmic}[1]
		\State Estimate ${\hat{\mathbf  R}}$ according to \eqref{eq:R1};
		\State Collect all anti-diagonal elements from ${\hat{\mathbf  R}}$ and construct  ${\bf R}_a$ in \eqref{eq:Ra_1};
		\State Calculate ${\bf R}_{a, \mathrm{DFT}}$ from ${\bf R}_a$ based on \eqref{eq:Ra_DFT};
		\State Search  $(\overline {i}_y,\overline{i}_z)= \arg \underset{({i}_y,{i}_z)}{{\max}}[{\bf R}_{\rm a,DFT}]_{{i}_y,{i}_z}$;
		\State Calculate AoAs $\hat u$ and $\hat v$ from \eqref{angle_initial};
		\State Get the refined AoAs by solving \eqref{eq:a_r_max};
		\State Estimate $\hat r$ by solving  \eqref{r_problem};
		\State Compute $\hat \beta$ according to \eqref{beta_e};
		\State Reconstruct ${\hat {\mathbf h}}$ according to \eqref{eq:h1} and \eqref{eq:b2}.
		\vspace{-0.1cm}
		
	\end{algorithmic}
\end{algorithm}%
Using the LS criterion, the distance estimation problem is expressed as
\begin{align}\label{r_problem}
\min\limits_{\hat{r}, \hat{\sigma}_{r}}\ \left\|\mathbf{f}(\hat u,\hat v) \times\left(\frac{1}{\hat r}+\hat \sigma_{r}\right)-\hat{\mathbf{a}}\right\|_{\rm F}^{2},
\end{align}
 where $\hat {r}$ and $\hat \sigma_{r}$ denote the estimate of $r$ and the corresponding estimate of parameter error, respectively. 
Define $\mathbf{B}$ as
\begin{align}
\mathbf{B}=\left(\begin{array}{cccc}
1 & 1 & \cdots & 1 \\
{[{\bf f}(\hat u, \hat v)]_0} & [{\bf f}(\hat u, \hat v)]_1 & \cdots & [{\bf f}(\hat u, \hat v)]_M
\end{array}\right)^{\rm T}.
\end{align}
 The LS solution of $r$ and $\sigma_{r}$ is denoted by
 \begin{align} \label{eq:d_esti}
 \left[\begin{array}{c}
 \hat{\sigma}_{r} \\
 \frac{1}{\hat{r}}
 \end{array}\right]=\left(\mathbf{B}^{\mathrm{T}} \mathbf{B}\right)^{-1} \mathbf{B}^{\mathrm{T}} \hat{\mathbf{a}}.
 \end{align}
After obtaining the distance estimate $\hat r$ in \eqref{eq:d_esti} and the estimated AoAs $\hat u$ and $ \hat v$ in \eqref{angle_fine}, the estimate of the complex channel gain $\beta$ is given as
\begin{align}\label{beta_e}
\hat{\beta}=\frac{1}{L} \sum_{l=1}^{L} \frac{\boldsymbol{b}(\hat u, \hat v, \hat r)^{\rm H} \boldsymbol{y}_{l}}{ \boldsymbol{b}(\hat u, \hat v, \hat r)^{\rm H} { \boldsymbol{b}(\hat u, \hat v, \hat r)}}.
\end{align}
Combing all the parameter estimates  in \eqref{angle_fine}, \eqref{eq:d_esti}, and \eqref{beta_e}, the original channel is ready to be reconstructed according to \eqref{eq:h1} and \eqref{eq:b2}.

The detailed steps of the proposed near-field channel estimation method are summarized in Algorithm 1. 
\vspace{-0.3cm}
\subsection {Complexity Analysis}
\begin{table}\scriptsize 
	\vspace{-0.3cm} 
	\caption{{\color{black} Complexity Comparison.} }
	\centering
	\label{table_time}
	\renewcommand{\arraystretch}{1.5}
	
	\begin{tabular}{|l|l|c|}  
		
		\hline  
		
		\makebox[0.05\textwidth][c]{\textbf{Method}}&\textbf{Computational Complexity}&\tabincell{c}{\textbf{Running Time}\\ \textbf{(Fig. 2) }}		 \\  
			
		\hline
		3D-MUSIC   & ${\mathcal O}(I_uI_vI_dM^2+M^3) $&$7.18\times 10^{7}$s    \\		
		\hline
		P-OMP in \cite{Dai_cb2} & ${\mathcal O}(S_uS_vS_dM+M^2) $&21.4080s    \\		
		\hline		
		NF-JCEL in \cite{Jin_arxiv} & \tabincell{l}{${\mathcal O} (I_uM^2M_{\rm Y}^2+I_dM^2)$} &11.8060s    \\

		\hline
		Proposed SADCE & \tabincell{l}{$\mathcal{O}(M^3+G_{\rm Y}G_{\rm Z}M)$} &1.9610s   \\
		\hline 
		
	\end{tabular}
\end{table}
The computation complexity of the proposed SADCE algorithm is dominated by three sequential steps: 1) the estimate of array covariance matrix; 2) the estimate of angle parameters; 3) the estimate of distance. 
Concretely, the complexity of step 1 in \eqref{eq:h_LS} and \eqref{eq:R1} amounts to $\mathcal{O}(M^2+ML)$. 
For the estimation of angle parameters in step 2, the dominant complexity corresponds to calculating the DFT transformation in \eqref{eq:Ra_DFT}, that is the  computational complexity of $\mathcal{O}(M{\rm log}(M))$, and to implementing the angle rotation in \eqref{eq:a_r_max} with the computational complexity of ${\mathcal O}(G_{\rm Y}G_{\rm Z}M)$, where $G_{\rm Z}$ and $G_{\rm Y}$ denote the search grids for $\Delta_{u}\in[-\frac{\lambda\pi}{dM_{\rm Z}},\frac{\lambda\pi}{dM_{\rm Z}}]$ and $\Delta_{v}\in[-\frac{\lambda\pi}{dM_{\rm Y}},\frac{\lambda\pi}{dM_{\rm Y}}]$, respectively. 
For step 3, the computation complexity is dominated by the matrix multiplications and matrix inversion with the computational complexity of ${\mathcal O}(M^3+M)$. 
Thus, the overall complexity of the proposed  algorithm is given by $\mathcal{O}(M^3+G_{\rm Y}G_{\rm Z}M)$.

For comparison, Table I summarizes the computational complexity of the proposed SADCE method and existing methods, including the 3D-MUSIC method, the near-field joint channel estimation and localization (NF-JCEL) algorithm in \cite{Jin_arxiv}, and the P-OMP method in \cite{Dai_cb2}, where $I_u$, $I_u$, and $I_d$ denote the numbers of searching grids for vertical angle, azimuth angle, and distance in 3D-MUSIC and \cite{Jin_arxiv}.
Constants $S_u$, $S_v$, and $S_d$ denote the number of codewords for sampled vertical angle, sampled azimuth angle, and sampled distance rings in  \cite{Dai_cb2}, respectively. Due to the fact that the proposed method narrows the searching range greatly, $\{G_{\rm Y},G_{\rm Z}\}$ usually satisfies $\{G_{\rm Y},G_{\rm Z}\}\ll \{S_u,S_v,I_u\}$. It is observed from Table I that the proposed method achieves a significant reduction in computational complexity.

\vspace{-0.3cm}
\section{Simulation Results}

\begin{figure}[t]
	\centering
	\vspace{-0.8cm}
	\subfloat[{\color{black}RMSE of AoA ($u$)}]{\includegraphics[width = 0.26\textwidth]{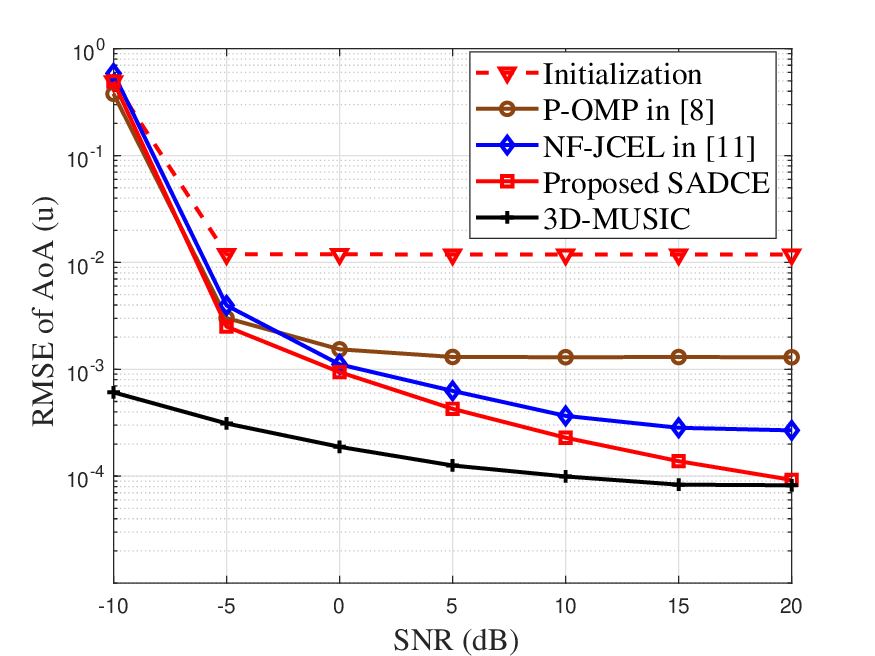}}
	\subfloat[{\color{black}RMSE of AoA ($v$)}]{\includegraphics[width = 0.26\textwidth]{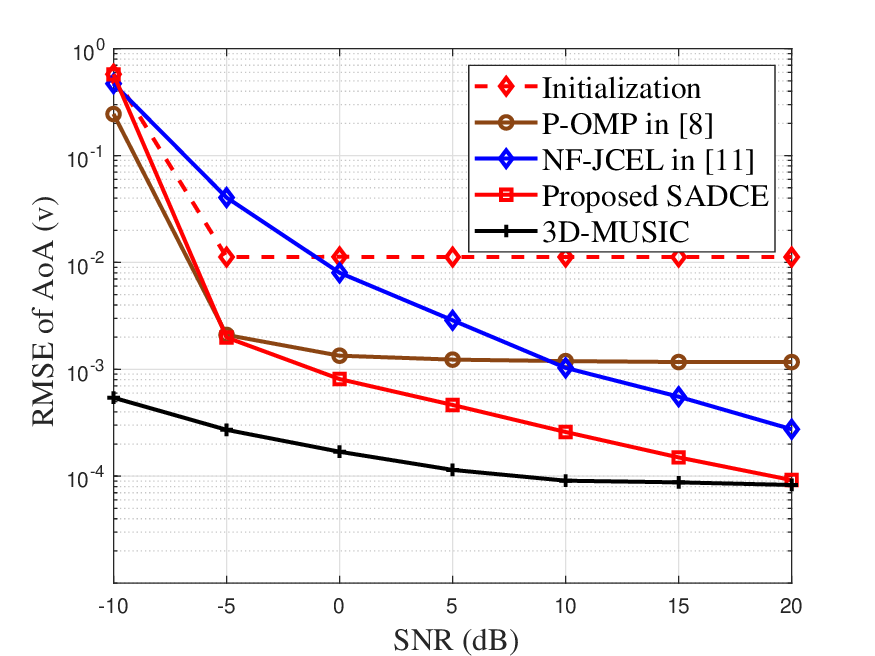}}
	\newline
	\vspace{-0.4cm}
	\subfloat[{\color{black}RMSE of distance ($r$)}]{\includegraphics[width = 0.26\textwidth]{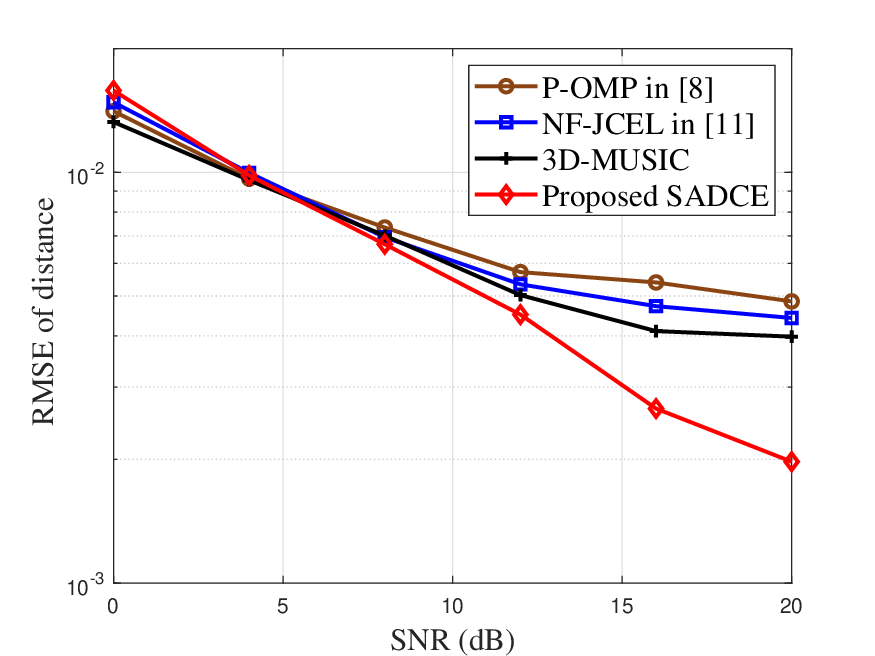}}
	\subfloat[{\color{black}NMSE of channel ($\bf h$)}]{\includegraphics[width = 0.26\textwidth]{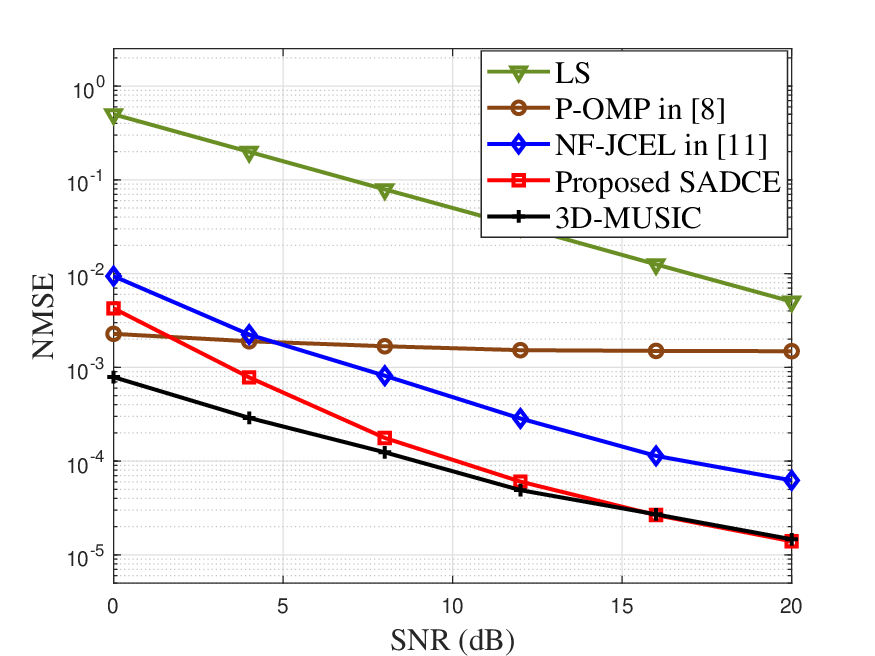}}
	\label{fig:label}
	\caption{\color{black}The average estimation performance versus  SNR.}
	\vspace{-0.5cm}
\end{figure}

In this section, we provide simulation results to validate the effectiveness of the proposed low-complexity SADCE method. The comparison is evaluated in terms of channel estimation and parameter recovery. {\color{black}For comparison, we choose the state-of-the-art techniques as benchmarks, i.e., the P-OMP method in \cite{Dai_cb2}, the NF-JCEL method in \cite{Jin_arxiv}, and the traditional methods, i.e., the LS method in \eqref{eq:h_LS} and the 3D-MUSIC method.} 
{\color{black} To better illustrate the effectiveness of angle compensation in Section III.A, the performance of the estimated AoAs in \eqref{angle_initial} are also provided in Fig.2 (a) and Fig.2 (b), which are labeled as ``Initialization''.} 

In the following simulations, we consider a scenario where users are distributed in a 5 m $\times $ 5 m square around (1.3, 0, 5) meters. 
The AP is installed on the wall at (1.3, 0, 6) meters. The complexity channel gain $\beta\sim{\mathcal{CN}}(0,1)$. The UPA at AP has $M = 41 \times 41$ antenna elements. 
For convenience, the pilot power $P$ is normalized to 1 and the signal-to-noise ratio (SNR) is calculated as $10{\rm log}_{10}(1/{\sigma^2})$ dB.  

Fig. 2 (a) and Fig. 2 (b) depict the root mean squared error (RMSE) performance of angle parameters as functions of SNR. It can be seen that the proposed SADCE method outperforms the other methods in angle estimation. 
Fig. 2 (c) illustrates the RMSE performance of distance between the user and the BS. In low SNR region, the distance estimation performance of the proposed algorithm is slightly worse than the NF-JCEL algorithm and P-OMP algorithm. Fortunately, the estimation accuracy rapidly improves with increasing SNR, and the distance estimation performance of the proposed method surpasses that of the other algorithms in the high SNR region.
 {\color{black}The reason is that the proposed close-form distance estimation method can break through the gridding limitation and achieve higher estimation accuracy than the existing methods.
	Fig. 2 (d) further evaluates the channel estimation performance. As can be seen in Fig. 2 (d), compared with the conventional 3D-MUSIC method, the proposed approach can approximate its performance with much lower computational complexity, achieving a more effective trade-off between accuracy and complexity than the existing state-of-the-art algorithms. 
}

\begin{figure}[t]
	\centering
	\vspace{-0.5cm}
	\includegraphics[width =2.1in]{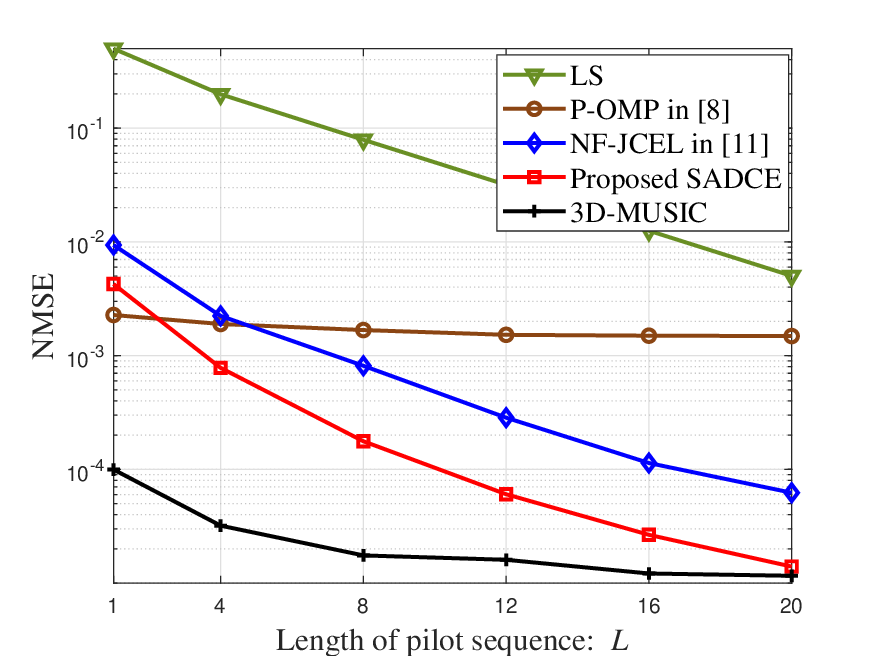}
	\caption{\color{black}The average NMSE of the estimated channel.}
	\label{Fig:P_pilot}
	\vspace{-0.5cm}
	
\end{figure}

{\color{black}
	Fig. \ref{Fig:P_pilot} depicts the impact of the length of pilots on the NMSE performance. It shows that the proposed SADCE method demonstrates a rapid improvement in channel estimation accuracy with increasing pilots. 
	For different values of the length of pilots, the proposed method consistently exhibits better estimation performance. It also indicates that the proposed method can effectively reduce the pilot overhead for the near-field estimation in XL-MIMO systems.
}
%

\vspace{-0.3cm}
\section{Conclusion}
\vspace{-0.1cm}
{\color{black}This paper studies the channel estimation problem for the XL-MIMO systems in the near field and proposes a low-complexity channel estimation method. The proposed SADCE method decouples the angle parameters and distance parameters, enabling low-complexity sequential estimation of these parameters. 
Simulation results verify that the proposed method achieves a significant improvement in computational efficiency}. 

\appendices
\vspace{-0.5cm}

\end{document}